\documentclass[12pt]{article}

\usepackage{url}

\usepackage{array}

\usepackage{booktabs}    

\usepackage{amsmath, amssymb, amsthm, graphicx}
\usepackage{natbib}
\usepackage{xcolor}
\usepackage{float}

\usepackage[margin=1.0in]{geometry}

\usepackage{hyperref}
\hypersetup{
  colorlinks   = true,    
  urlcolor     = blue,    
  linkcolor    = blue,    
  citecolor    = blue     
}

\def\thesection{\arabic{section}}
\def\bE{\mathbb{E}} 

\def\vec#1{\mathchoice{\mbox{\boldmath$\displaystyle\bf#1$}}
{\mbox{\boldmath$\textstyle\bf#1$}}
{\mbox{\boldmath$\scriptstyle\bf#1$}}
{\mbox{\boldmath$\scriptscriptstyle\bf#1$}}}
\newtheorem{thm}{Theorem}[section]

 \newtheorem{lem}[thm]{Lemma}

 \numberwithin{equation}{section}
 
\usepackage{authblk}
\title{Estimation of Group Means in Generalized Linear Mixed Models}
\author[1]{Jiexin Duan\thanks{PhD candidate, Department of Statistics, Purdue University, West Lafayette, IN 47906.}}
\author[1]{Michael Levine\thanks{Corresponding Author. Associate Professor, Department of Statistics, Purdue University, West Lafayette, IN 47906. (Email:mlevins@purdue.edu). }}
\author[2]{Junxiang Luo\thanks{Biostatistics and programming, Sanofi, 55 Corporate Drive
Bridgewater, NJ 08807, USA.}}
\author[3]{Yongming Qu\thanks{Department of Biometrics, Eli Lilly Corporate Center, Drop Code 2232, Indianapolis, IN, USA. }}

\affil[1]{Department of Statistics, Purdue University}
\affil[2]{Biostatistics and programming, Sanofi}
\affil[3]{Department of Biometrics, Eli Lilly} 

\date{}

\begin{document}
\maketitle


\begin{abstract}
In this manuscript, we investigate the concept of the mean response for a treatment group mean as well as its estimation and prediction for generalized linear models with a subject-wise random effect.  Generalized linear models are commonly used to analyze categorical data. The model-based mean for a treatment group usually estimates the response at the mean covariate. However, the mean response for the treatment group for studied population is at least equally important in the context of clinical trials. New methods were proposed to estimate such a mean response in generalized linear models; however, this has only been done when there are no random effects in the model.  We suggest that, in a generalized linear mixed model (GLMM), there are at least two possible definitions of a treatment group mean response that can serve as estimation/prediction targets. The estimation of these treatment group means is important for healthcare professionals to be able to understand the absolute benefit versus risk. For both of these treatment group means, we propose a new set of methods that suggests how to estimate/predict both of them in a GLMM models with a univariate subject-wise random effect. Our methods also suggest an easy way of constructing corresponding confidence and prediction intervals for both possible treatment group means. Simulations show that proposed confidence and prediction intervals provide correct empirical coverage probability under most circumstances.  Proposed methods have also been applied to analyze hypoglycemia data from diabetes clinical trials.
\end{abstract}


\newpage

\section{Introduction}

Generalized linear mixed models (GLMM) are used today rather often for analysis of non-normal data with random effects. These data are commonly encountered in medical research and across many disciplines. They are particularly convenient for the analysis of categorical, e.g. binomial, Poisson, negative binomial etc. data. An excellent introduction to GLMM's can be found in \cite{demidenko2013mixed} and \cite{jiang2007linear}. 


The research described in this manuscript has been motivated by modeling of clinical trials, especially diabetes clinical trials. In that area, it has been rather common historically to focus on the estimation of the treatment effect. While this is often important for sponsors, regulatory agencies, and the medical community at large, the estimation of treatment group means is also important for healthcare professionals to understand the absolute benefit versus risk. Thus, it is important to identify the appropriate methods to provide good estimators for the group means. Most types of statistical software, such as SAS or R, provide least squares group means. These means are estimated at the mean value of the baseline covariates; of course, for basic linear models, these means are equivalent to the average of individual response estimates for the population studied. Moreover, in this case, the least squares group means are consistent estimators of the true group means. For generalized linear models (GLM), the average of individual response estimates for the given population is not the same as the group mean estimated at the mean value of baseline covariates; moreover, as \cite{qu2015estimation} pointed out, the latter is not even a consistent estimator of the true group mean. Estimating the average of individual responses in a GLM has been publicly available for practitioners as a part of Stata software margins command; see e.g. \cite{statacorp2013stata}. \cite{qu2015estimation} addressed the issue of estimating the average of individual response estimates for GLM - type models in the randomized clinical trial context. They described a possible way of estimating the mean outcome assuming assignment to a particular treatment for each patient in the trial, using each patient's observed baseline covariates, and averaging these estimates over all patients in the trial.
Additional research in this direction has been conducted by \cite{bartlett2018covariate}. They introduced a new semiparametric estimator of the group mean that is consistent even if the working model is misspecified.  Both simple and stratified randomization schemes have also been discussed in detail.

The model used by \cite{qu2015estimation} and \cite{bartlett2018covariate} has a shortcoming in that it does not allow for a subject-wise random effect. As an example, such an effect may be needed to model dependence of observations for the same subject over time when time is present as a covariate. Doing so in this context changes the model considered in \cite{qu2015estimation} and \cite{bartlett2018covariate} from GLM to a GLMM. The estimation of treatment group means for this new model is as important as before and, to the best of our knowledge, has hardly been addressed in the literature so far.  

Due to the presence of random effects in a GLMM, the average response over a treatment group can be characterized in several different ways. In this manuscript we suggest two different ways of characterizing and estimating treatment group means in GLMM in the randomized trial context. The first approach uses the so-called ``marginal means" (using the terminology of \cite{bartlett2018covariate}) that are averaged over random effects to characterize treatment group means. We propose some natural estimators for these marginal means and a possible way of constructing confidence intervals for them. All of this is described in detail in the Section \eqref{conf_ref}. Our approach is frequentist in nature; a Bayesian approach has been attempted in \cite{LalQu}. The second approach uses conditional (on random effects) means to characterize treatment group means. As in the case of the first approach, we also provide some inferential procedures; in particular, we construct several versions of possible ``prediction" intervals for the unknown conditional means. The term ``prediction" interval is used here intentionally due to the fact that these intervals are conditional on the random effect. All of this is done in Section \eqref{pred_ref}. Section \eqref{sim} describes several simulation studies that illustrate our results. Finally, Section \eqref{sum} shows how proposed methods can be applied to a real dataset.  

\section{Characterization of the group effect with the marginal group mean}\label{conf_ref}

We begin with the detailed description of our model. Let $y_{ij}$ be the $j$th response of $i$th subject, $j=1,\ldots, n_{i},$ $i=1,\ldots,K,$ and $b_{i}\sim N(0,\sigma^{2})$ is the subject-wise random effect. Random effects are assumed to be independent between subjects. In this setting, the total number of observations is $N=\sum_{i=1}^{K}n_{i}.$ For each subject, let $\vec \beta$ be a $p\times 1$ vector of fixed effects. Usually, a $p\times 1$ covariate vector is denoted $\vec x_{ij}.$ Then, we assume that the responses are generated by a generalized linear model conditional on the random effects with a linear predictors $\eta_{ij}=\vec x_{ij}^{'}\vec\beta+b_{i}$ where the mean $\mu_{ij}$ is connected to the linear predictor $\eta_{ij}$ through the canonical link function $g$ as $g(\mu_{ij})=\eta_{ij}.$ Following \cite{booth1998standard}, we will use $\vec x_{i}$ as a generic vector of fixed covariate values that may or may not be equal to $\vec x_{ij}$ and, therefore, will view our generalized linear mixed model as 
\begin{equation}\label{model1}
\eta_{i}=\vec x_{i}^{'}\vec\beta+b_{i}.
\end{equation}
For each subject, the corresponding marginal mean is, then, $\mu_{i}=\bE\,g^{-1}(\eta_{i})=\bE\,g^{-1}(\vec{x}_{i}^{'}\vec \beta+b_{i})$ where the expectation is with respect to the distribution of the random component. 
 
Typically, covariate vectors consist of the treatment assignment, some additional covariate, e.g. time or gender, and the baseline variable. An example of a baseline variable may be some important prognostic factor, e.g. whether to use sulfonylurea as a background therapy during the trial of an anti-diabetes medication. We will also be viewing observations as belonging to groups where each group is defined by its value of the treatment factor level and its value of the additional covariate (e.g. the time point or a gender category.) We assume that, in total, there are $Q$ groups and each group consists of $N_{q}$ observations, $q=1,\ldots,Q.$ Of course, the total number of observations can also be expressed as $N=\sum_{q=1}^{Q}N_{q}.$ The same notation $q$ will also be used to denote a set of indices describing observations/subjects in the $q$th group. 
 
There are several ways to define the group mean in the model \eqref{model1}. The first one that we consider in the current section, is the group average of individual marginal means for each subjects. In other words, we define the marginal group mean as 
\begin{equation}\label{true_gr_mean}
   \mu_{q}=\frac{1}{N_{q}}\sum_{i\in q} \bE\,g^{-1}(\eta_{i}). 
\end{equation}  Such a mean is a fixed parameter and, therefore, in addition to the point estimate, we also suggest possible confidence intervals for it.

In order to do so, we have to be able to estimate parameters of the GLMM, namely $\vec \beta$ and $\sigma^{2}$. 
We obtain estimated fixed effect parameters $\vec{\hat \beta}$ and an estimated variance $\hat{\sigma}^{2}$ by maximizing the marginal likelihood $l(\vec \beta,\sigma^{2})$. The general form of the marginal likelihood is given, for convenience, in the Appendix Section \eqref{marglik}. It is rather difficult, in general, to evaluate the marginal likelihood for a GLMM. Two main approximation methods that are used in practice are either a pseudo-likelihood approach (see e.g. \cite{schall1991estimation}) or an integral approximation that uses the Gauss-Hermit quadrature or Laplace approximation (see e.g. \cite{pinheiro1995approximations}). Pseudo-likelihood does not work well for a logistic mixed model when the number of observations for the majority of subjects (locations) is small; moreover, it is also known to perform badly in situations where the distribution of the data comes from a family with more than one parameter (e.g. negative binomial). For a more detailed discussion of this problem see e.g. \cite{stroup2016generalized} p.142. Since both logistic and negative binomial models are of great importance  in clinical trial applications, we will use the integral approximation approach to computation of the marginal likelihood $l(\vec \beta,\sigma^{2}).$

Finally, when estimators of the model parameters are available, a natural estimator of the group mean can be defined as 
\begin{equation}\label{gr_mean}
\hat \mu_{q}=\frac{1}{N_{q}}\sum_{i \in q}\hat \mu_{i}
\end{equation}
where $\hat \mu_{i}$ is the estimate of $\bE\,g^{-1}(\eta_{i})=\bE\,g^{-1}(\vec{x}_{i}^{'}{\vec \beta}+{b}_{i}).$ In the future, we will also call this estimator an {\it unconditional} or a {\it marginal} group mean. Now, we are going to consider two separate cases: the case of a logistic regression with a random intercept and the case of the negative binomial regression with a random intercept.

\subsection{Logistic case}\label{logistic}
As a first scenario, we assume that the data is binary and that the model is a logistic regression with a Gaussian random effect. As a first step, we start with estimation of the marginal group mean. This is, however, not a straightforward task because $\mu_{i}=\bE\,g^{-1}(\vec x_{i}^{'}\beta+b_{i})$ cannot be written down as a closed form expression as a function of parameters $\vec\beta$ and $\sigma^{2}.$ Indeed, since the link function $g$ is a logistic one, we have $g^{-1}(\vec{x}_{i}^{'}\vec \beta+b_{i})=\frac{\exp(\vec{x}_{i}^{'}\vec \beta+b_{i})}{1+\exp(\vec{x}_{i}^{'}\vec \beta+b_{i})}$. The expectation of the above can be written down as
\begin{equation}\label{log.norm}
\int \frac{\exp(\vec{x}_{i}^{'}\vec \beta+b_{i})}{1+\exp(\vec{x}_{i}^{'}\vec \beta+b_{i})}\frac{1}{\sqrt{2\pi\sigma^{2}}}\exp{\left(-\frac{1}{2\sigma^2}b_{i}^{2}\right)}\,db_{i}
\end{equation}
and is known as the logistic-normal integral. This integral cannot be obtained in closed form, and some approximation is necessary. A common approximation used to compute it can be found in \cite{zeger1988models}:
\begin{equation}\label{zeger}
\mu_{i}\approx \frac{\exp(c\vec{x}_{i}^{'}\vec \beta)}{1+\exp(c\vec{x}_{i}^{'}\vec \beta)}.
\end{equation}
where the constant $c$ in \eqref{zeger} is approximately equal to $[1+0.346\sigma^{2}]^{-1/2}$.  With the above so-called Zeger's approximation in mind, we suggest estimating $\mu_{i}$ using a plug-in estimator $\hat \mu_{i}=\frac{\exp(\hat c \vec{x}_{i}^{'}\hat {\vec \beta})}{1+\exp(\hat c \vec{x}_{i}^{'}\hat {\vec \beta})}$ where $\hat c=[1+0.346\hat \sigma^{2}]^{-1/2}$. 

In order to construct a confidence interval for the marginal group mean $\mu_{q}$, we need to estimate the variance of the estimated mean $\hat \mu_{q}.$ Due to \eqref{gr_mean}, the variance of the estimated group mean response for $q$th group is  
\begin{align}\label{var}
&Var\,(\hat\mu_{q})=\frac{1}{N_{q}^{2}}\sum_{i\in q}Var\,(\hat \mu_{i})\\
&+\frac{2}{N_{q}^{2}}\sum_{\substack{i_{1}<i_{2}\\ i_{1},i_{2}\in q}} Cov \left(\hat \mu_{i_{1}}, \hat \mu_{i_{2}}\right)\nonumber.
\end{align}
This suggests that it is necessary to estimate variance of individual estimated means $\hat\mu_{i}$ and the corresponding covariances $Cov \left(\hat \mu_{i_{1}}, \hat \mu_{i_{2}}\right),$ in order to be able to construct a confidence interval for $\mu_{q}$. The details of estimation procedures can be found in the Appendix Section \eqref{margmean}. 

Now that we can estimate the variance of the group mean $\hat \mu_{q}$, we can also construct a confidence interval for it. Recall that $\hat \mu_{q}=\frac{1}{N_{q}}\sum_{i \in q}\hat\mu_{i}$ where each $\hat \mu_{i}\approx \frac{\exp{(\hat c\vec{x}_{i}^{'}\hat{\vec \beta})}}{1+\exp{(\hat c\vec{x}_{i}^{'}\hat{\vec \beta})}}$. We suggest using an approximate large sample $100(1-\alpha)\%$ confidence interval of the form  
\begin{equation}\label{CI_log1}
\hat \mu_{q}\pm z_{1-\alpha/2}\sqrt{Var(\hat \mu_{q})} 
\end{equation}
that should be used for sufficiently large group size $N_{q}$. Given that individual random variables $\hat\mu_{i}$ are bounded and, therefore, have the finite third moment, it is possible to show that the rate of convergence in the corresponding central limit theorem for sums of $\hat\mu_{i}$ will have the standard square root rate see e.g. \cite{rio1996theoreme}. Because of this, and also because variances of $\hat\mu_{i}$ are bounded for any $i,$ we suggest that the confidence interval \eqref{CI_log1} can be used in practice already for relatively small group sizes of about $40$ to $50$ in the same way as is done for standard sequences of independent random variables with finite variances. In the future, we will refer to this confidence interval as the direct interval. For comparison purposes, we will also consider another confidence interval that is based on the use of delta-method to compute the variance of the logit of $\hat\mu_{q}.$ More specifically, it is not hard to see that, by the delta method,  
\[
Var \left[\log\left(\frac{\hat\mu_{q}}{1-\hat\mu_{q}}\right)\right]=\frac{Var(\hat\mu_{q})}{[\hat\mu_{q}(1-\hat\mu_{q})]^{2}}.
\]
Next, we construct a $100(1-\alpha)\%$ central limit theorem based interval for the $\log\left(\frac{\mu_{q}}{1-\mu_{q}}\right)$ and apply the inverse logit transformation to its end points. The final expression for such an interval is $(L(\hat\mu_{q}),U(\hat\mu_{q}))$ where
the lower bound 
\begin{equation}\label{lbound}
L(\hat\mu_{q})=\frac{\hat\mu_{q}/(1-\hat\mu_{q})\exp{\left(-\frac{z_{1-\alpha/2}\sqrt{Var(\hat\mu_{q})}}{\hat\mu_{q}(1-\hat\mu_{q})}\right)}}{1+\hat\mu_{q}/(1-\hat\mu_{q})\exp{\left(-\frac{z_{1-\alpha/2}\sqrt{Var(\hat\mu_{q})}}{\hat\mu_{q}(1-\hat\mu_{q})}\right)}}
\end{equation}
and the upper bound 
\begin{equation}\label{ubound}
U(\hat\mu_{q})=\frac{\hat\mu_{q}/(1-\hat\mu_{q})\exp{\left(\frac{z_{1-\alpha/2}\sqrt{Var(\hat\mu_{q})}}{\hat\mu_{q}(1-\hat\mu_{q})}\right)}}{1+\hat\mu_{q}/(1-\hat\mu_{q})\exp{\left(\frac{z_{1-\alpha/2}\sqrt{Var(\hat\mu_{q})}}{\hat\mu_{q}(1-\hat\mu_{q})}\right)}}. 
\end{equation}
The details can be found in the Appendix of \cite{qu2015estimation}. In what follows, we will refer to this interval as the inverse confidence interval.

\subsection{Negative binomial case}\label{neg.bin}

In this section we assume that the response data $Y$ have a negative binomial distribution with the mean parameter $\mu$ and the so-called size $\kappa$. This implies that the mean of the distribution is $\bE\,Y=\mu$ and the variance is $Var(Y)=\mu+\frac{\mu^{2}}{\kappa}$. This parameterization results from modeling an overdispersion in a Poisson distribution with the Poisson mean being a gamma random variable with the mean $\mu$ and the constant index $\kappa$. For more details see e.g. \cite{mccullagh2018generalized} p.198-199. In the negative binomial case, the canonical link function is the natural log; thus, the inverse link function is the exponent and so $\mu_{i}=\bE\,\exp(\vec{x}_{i}^{'}\vec \beta+b_{i})=\exp(\vec{x}_{i}^{'}\vec \beta+\sigma^{2}/2)$. Unlike in the logistic case, here we obtain a closed form expression for $\mu_{i}$. To keep notation clear, we denote the estimated mean $\hat \mu_{i}\equiv \mu_{i}(\hat \beta,\hat \sigma)=\exp(\vec{x}_{i}^{'}\hat{\vec \beta}+\hat\sigma^{2}/2).$  We also denote $\nu_{i}=\log \mu_{i}$ and $\hat \nu_{i}=\log \hat\mu_{i}.$ We denote the vector of parameters $\psi=(\vec\beta,\sigma^{2})^{'}$ and the corresponding vector of estimators $\hat\psi=(\hat{\vec\beta},\hat{\sigma}^{2})^{'}.$ Finally, let $\Sigma_{\hat\psi}$ be the covariance matrix of the vector $\hat\psi.$ In order to obtain a confidence interval for $\mu_{q}$, we need to provide a suitable approximation of the variance of 
\begin{equation}\label{sum_nb}
\hat \mu_{q}=\frac{1}{N_{q}}\sum_{i \in q}\hat \mu_{i}.
\end{equation} 
To do so, we will use the following empirical estimator of the variance of $\hat\mu_{q}:$ 
\begin{align}\label{varmu}
&Var\,\hat\mu_{q}=\sum_{i_{1}\in q}\sum_{i_{2}\in q}\exp \left(\hat\nu_{i_{1}}+\hat\nu_{i_{2}}+\frac{1}{2}\left(\hat\sigma_{i_{1}}^{2}+\hat\sigma_{i_{2}}^{2}\right)\right)\\
&\times \left(\exp {\hat\sigma^{2}_{i_{1},i_{2}}}-1\right)\nonumber
\end{align}

where $\hat\sigma^{2}_{i}:=Var\,\hat\nu_{i}\approx \nabla_{\hat\psi}^{'}\hat\nu_{i}\Sigma_{\hat\psi}\nabla_{\hat\psi}\hat\nu_{i},$ and $i\in (i_{1},i_{2}),$ and $\hat\sigma^{2}_{i_{1},i_{2}}:=Cov\,(\hat\nu_{i_{1}},\hat\nu_{i_{2}})\approx \nabla_{\hat\psi}^{'}\hat\nu_{i_{1}}\Sigma_{\hat\psi}\nabla_{\hat\psi}\hat\nu_{i_{2}}.$ See the Appendix Section \eqref{negbin} for the detailed discussion of this approximation.

Now we can obtain, similarly to \eqref{CI_log1}, a straightforward large sample confidence interval for the true $\mu_{q}$ based on the use of CLT applied to the sum in \eqref{sum_nb}. Using the approximate variance of $\hat\mu_{q}$ from \eqref{varmu}, such a large sample $100(1-\alpha)\%$ confidence interval will take the form  \begin{equation}\label{CI_negb1}
\hat \mu_{q}\pm z_{1-\alpha/2}\sqrt{Var(\hat \mu_{q})} 
\end{equation}
that is approximately correct for sufficiently large group size $N_{q}.$  In the future, we will also refer to \eqref{CI_negb1} as a direct interval. For comparison purposes, we will consider two other  confidence intervals as well. The first of these is analogous to the inverse confidence interval introduced earlier for the logistic data case except that the link function is now a log function (and the inverse link is the exponential function). We will refer to this interval as the inverse interval, as before. Let $LNB(\hat\mu_{q})=\hat\mu_{q}\exp{\left(-z_{1-\alpha/2}\sqrt{Var(\hat\mu_{q})}/\hat\mu_{q}\right)},$   $UNB(\hat\mu_{q})=\hat\mu_{q}\exp{\left(z_{1-\alpha/2}\sqrt{Var(\hat\mu_{q})}/\hat\mu_{q}\right)}.$ Then, the final form of this interval is
\begin{equation}\label{inv.negb}
\left(LNB(\hat\mu_{q})\,,UNB(\hat\mu_{q})\right)
\end{equation}
Finally, the third confidence interval that we propose is based on the idea of approximating a sum of lognormally distributed random variables. For simplicity, we will call this one a lognormal confidence interval. First, recall that $N_{q}\hat\mu_{q}=\sum_{i\in q}\hat \mu_{i}$. Next, we recall that each estimated $\hat \mu_{i}=\exp(\vec{x}_{i}^{'}\hat{\vec \beta}+\hat\sigma^{2}/2)$ and that individual estimated parameters $\hat{\vec \beta}$ and $\hat\sigma^{2}$ are approximately normally distributed as remarked earlier. Thus, we can assume that $\hat \mu_{i}$ has an approximate lognormal distribution. Thus, it seems possible to approximate the distribution of $N_{q}\hat\mu_{q}$ if some approximation for the distribution of a sum of lognormal random variables is available. 

A question of approximating a sum of lognormal random variables acquired substantial importance first in studies of cellular radio systems because the interference with a broadcast can typically be modeled as a sum of (usually dependent) lognormal random variables. This sum does not have a closed form distribution; hence, an approximation is required. Until recently, Schwartz-Yeh method \cite{schwartz1982distribution}, which is iterative in nature, has been the most used approximation method. An alternative is an extended version of Fenton and Wilkinson methods; see, for example, \cite{safak1994moments} and \cite{pratesi2000outage}. Both of these methods are based on the idea of approximating the sum of lognormal distributions by another lognormal distribution.  We will use just this basic idea in our approach. We already noted that $N_{q}\hat\mu_{q}=\sum_{i\in q}\hat \mu_{i}$ can be viewed approximately as a sum of lognormal random variables and so we will approximate it as yet another lognormal random variable with the mean and the variance equal to the mean and the variance of the sum of $\hat\mu_{i}.$ Thus, if $\hat\mu_{i}$ is assumed to be lognormally distributed, then for any positive $N_{q}$, $N_{q}\hat \mu_{q}$ is also approximately lognormally distributed with the mean equal to $N_{q}\mu_{q}=N_{q}\sum_{i\in q}\exp(\vec{x}_{i}^{'}\vec \beta+\sigma^{2}/2)$ and the variance $N_{q}^{2}Var(\hat \mu_{q})$. 

Let the significance level of the interval to be $\alpha.$ Then, let us denote $\xi_{L}$ and $\xi_{U}$ $\alpha/2$ and $1-\alpha/2$ percentiles of the lognormal distribution with the mean $N_{q}\sum_{i\in q}\exp(\vec{x}_{i}^{'}\vec \beta+\sigma^{2}/2)$ and the variance $N_{q}^{2}Var(\hat \mu_{q}).$  Then, the lower and upper bounds of the $100(1-\alpha)\%$ confidence interval for $\hat\mu_{q}$ are 
\begin{equation}\label{lognormal.negb}
\left[\frac{\xi_{L}}{N_{q}} , \frac{\xi_{U}}{N_{q}}\right].
\end{equation}

\section{Characterization of the group effect with the conditional group mean}\label{pred_ref}

As an alternative to the marginal group mean, we also consider another possible way of characterizing the group mean in a GLMM with a random effect. To introduce the idea, we begin with a brief restatement of the setting in which we work. First, recall that observations are recorded as $y_{ij}$ -  $j$th response of $i$th subject, $j=1,\ldots,n_{i},$ $i=1,\ldots,K,$ and $b_{i}\sim N(0,\sigma^{2})$ is the subject-wise random effect. Random effects are assumed independent between subjects. The responses come from a generalized linear model (conditional on the random effects) with linear predictors
\begin{equation}\label{pred}
\eta_{ij}=\vec x_{ij}^{'}\vec\beta+b_{i};
\end{equation}
here, $\vec\beta$ is a $p\times 1$ vector of fixed effects, and $\vec x_{ij}$ is a $p\times 1$ vector of covariates.
From the practical viewpoint, it may be more convenient to use the following framework. Let us define a vector of covariate values for a generic $i$th subject as $\vec z_{i}$ of dimensionality $K\times 1$ with the $i$th element being equal to $1$ and all others to zero. Let also $\vec x_{i}$ be a generic vector of fixed covariate values that may or may not be equal to $\vec x_{ij}.$  Finally, we can view all of the random effects taken jointly as a vector $\vec b=(b_{1},\ldots,b_{K}).$ Our problem is the estimation of and inference about the group mean. This problem is, effectively, a problem of estimation and inference about linear combinations of the form
\begin{equation}\label{pred1}
\eta_{i}=\vec x_{i}^{'}\vec\beta+\vec z_{i}^{'}\vec b.
\end{equation}
As before, observations belong to one of $Q$ groups where each group consists of $N_{q}$ observations, $q=1,\ldots,Q.$  Of course, the total number of observations can also be expressed as $N=\sum_{q=1}^{Q}N_{q}.$ We also use $q$ to denote a set of indices describing the subjects in the $q$th group. We will now measure the contribution of each subject in a given group to the overall group mean by a conditional mean $\lambda_{i}=\bE\,(y_{i}|b_{i}).$ This mean is connected to the linear predictor $\eta_{i}$ through the canonical link function $g$ as $g(\lambda_{i})=\eta_{i}.$ For $q$th group, we define the conditional group mean 
\begin{equation}\label{gr.mean}
\lambda_q=\frac{1}{N_{q}}\sum_{i\in q}\lambda_{i} =\frac{1}{N_{q}}\sum_{i\in q}g^{-1}(\eta_{i})
\end{equation}
Our eventual purpose will be to predict the group mean \eqref{gr.mean} and to construct a prediction interval for such a predicted group mean. Note that we are talking about prediction, rather than estimation, since the mean \eqref{gr.mean} is a random variable. 

To suggest a good predictor for $\lambda$ we start, first, with the distribution of responses. Recall that the responses are (conditionally) independent and their conditional density function is of the form 
\begin{equation}\label{cond.distr}
f(y_{ij}|b_{i},\vec\beta,\sigma_{0}^{2})=\exp\left\{\frac{w_{ij}}{\sigma_{0}^{2}}(y_{ij}\theta_{ij}-c(\theta_{ij}))+d(y_{ij},\sigma^{2}/w_{ij})\right\};
\end{equation}
in the above, $w_{ij}$ are known weights, $\sigma_{0}^{2}$ is the dispersion parameter, and canonical parameters are related because $\lambda_{ij}=\bE\,(y_{ij}|b_{i})=c^{'}(\theta_{ij})$ (see e.g. \cite{mccullagh2018generalized}). Let the vector of observations for $i$th subject be $\vec y_{i}=(y_{i1},\ldots,y_{in_{i}})^{'}.$ Let also $X_{i}$ be the corresponding $n_{i}\times p$ covariate matrix of fixed effects and $J_{i}$ be the $n_{i}\times 1$ vector of ones. Stacking row vectors $\vec z_{i}^{'},$ $i=1,\ldots,K$, on top of each other while repeating each one $n_{i}$ times produces a complete random covariate matrix $Z$ of dimensionality $N\times K.$ Let $V(\mu_{ij})=c^{''}(\theta_{ij})$ be the variance function for the generalized linear model defined in \eqref{pred} and \eqref{cond.distr}. The $n_{i}\times n_{i}$ diagonal matrix of iterative weights $\frac{w_{ij}}{\sigma_{0}^{2}V(\mu_{ij})[g^{'}(\mu_{ij})]^{2}},$ $j=1,\ldots, n_{i}$ for the $i$th subject is denoted $W_{i}.$ The complete data vector, covariate, and weight matrices are $\vec y(N\times 1),$ $X(N\times p),$ and $W(N\times N),$ respectively. It is assumed that the matrix $X$ has a full column rank. In addition, we let ${\cal G}(K\times K)$ be the diagonal matrix with $\sigma^{2}$ on its diagonal. Finally, we also denote the complete parameter vector of the model \eqref{pred}-\eqref{cond.distr} $\vec\psi=(\vec\beta^{'},\sigma_{0}^{2},\sigma^{2})^{'}$ and the subvector of variance components $\vec\sigma^{2}=(\sigma_{0}^{2},\sigma^{2})^{'}.$  

Any inference about random effects $b_{i}$ will be based on the conditional likelihood 
\begin{align}\label{llik}
&h(b_{i}|\vec y_{i};\vec \psi)=\frac{1}{L_{i}(\vec \psi;\vec y_{i})}\left\{\prod_{j=1}^{n_{i}}f(y_{ij};b_{i},\vec \beta,\sigma_{0}^{2})\right\}[\sqrt{2\pi}]^{-1}[\sigma^{2}]^{-n_{i}/2}\\
&\times \exp\left\{-\frac{b_{i}^{2}}{\sigma^{2}}\right\}\nonumber
\end{align}
where $L_{i}(\vec\psi;\vec y_{i})$ is the normalizing expression. Let 
\begin{equation*}
l_{i}=\sum_{j=1}^{n_{i}}\log f(y_{ij};b_{i},\vec\beta,\sigma_{0}^{2})-\log (\sqrt{2\pi})
-\frac{n_{i}}{2\log (\sigma^{2})}-\frac{b_{i}^{2}}{\sigma^{2}}
\end{equation*}
An alternative, and often convenient form for the conditional likelihood in \eqref{llik} is $h(\vec b_{i}|\vec y_{i};\vec \psi)=L_{i}^{-1}\exp{\{l_{i}(b_{i})\}}$ where, for convenience, the dependence on $\vec y_{i}$ and $\vec \psi$ has been suppressed. As is common in mixed model theory, the natural predictor of the random effect $b_{i}$ is its conditional mean so that the predicted $i$th random effect is defined as $b_{i}(\vec \psi;\vec y_{i}):=\bE_{\psi}\,(b_{i}|\vec y_{i}).$ In other words, only observations on $i$th subject, contained in the vector $\vec y_{i}$, are relevant for prediction of $b_{i}$. To make the notation easier, we will omit the parameter vector $\vec\psi$ in the subscript in the future, unless absolutely necessary. Similar approach is usually adopted in most of GLMM literature; see, e.g. \cite{booth1998standard}.

A natural point predictor for the vector $\vec b$ is its conditional mean, $\vec b(\vec\psi;\vec y):=\bE_{\vec\psi}(\vec b|\vec y).$ Clearly, the point predictor for $\eta_{i}$ is $\eta_{i}(\vec\psi;\vec y):=\vec x_{i}^{'}\vec\beta+\vec z_{i}^{'}\vec b(\vec\psi;\vec y).$ Therefore, in practice it makes sense to predict the group mean \eqref{gr.mean} with \begin{equation}\label{pred.gr.mean}\hat\lambda_q=\frac{1}{N_{q}}\sum_{i\in q}\hat\lambda_{i}=\frac{1}{N_{q}}\sum_{i\in q}g^{-1}(\eta_{i}(\vec{\hat\psi};\vec y))\end{equation} where $\hat{\vec \psi}$ is a consistent estimator of $\vec\psi.$ At this point, it remains to estimate a variance of \eqref{pred.gr.mean} and construct at least one possible prediction interval for it. In order to do so, however, we need a few additional definitions first. Let $\hat b_{i}$ denote the maximizer of $l_{i}(b_{i})$ that satisfies the equation $l_{i}^{'}(b_{i})=0.$ Let $\tilde W_{i}$ denote the matrix of iterative weights for the $i$th subject evaluated at $b_{i}=\hat b_{i}$ and $\tilde W$ the complete matrix of iterative weights for all subjects evaluated at $\hat b=(\hat b_{1},\ldots,\hat b_{K})^{'}.$ Next, let $\vec z_{i},$ be the random covariate vector for $i$th subject in $q$th group, $i=1,\ldots,N_{q}.$ Then, we can define $Z_{q}$ is the $K\times N_{q}$ matrix of these random covariate vectors of subjects in the $q$th group. Also, let $X_{q},$ similarly, be $p\times N_{q}$ matrix  of fixed covariate vectors of subjects in $q$th group. For the $q$th group, $X_{q}$ and $Z_{q}$, taken together, constitute a matrix $\begin{pmatrix} X_{q}, Z_{q}\end{pmatrix}$ that consists of $N_{q}$ columns; each column is of dimensionality $(p+K)\times 1.$ The first $p$ elements are fixed covariates of the $i$th subject in that group, $i=1,\ldots,N_{q}$ and the other $K$ elements are random covariates of the same subject. Finally, $J_{N_{q}}$ is a vector of $1$'s of dimensions $N_{q}\times 1.$ Now we are ready to state the main result of this Section. 

\begin{lem}\label{var_pred}
Let $\vec{\hat \eta}_{q}$ be the $N_{q}\times 1$ vector of $\eta_{i}(\vec{\hat\psi};\vec y).$ The variance of the predicted group mean $\hat\lambda_{q}$ can be approximately computed as 
\begin{align}\label{pr.var}
&Var (\hat \lambda_{q})=Var \left( \frac{1}{N_{q}}J_{N_{q}}^{'}g^{-1}(\hat{\vec\eta}_{q})\right)\\
&=\frac{1}{N_{q}^{2}}J_{N_{q}}^{'} \{\nabla g^{-1}(\vec {\hat\eta}_{q})\}^{'}(C(\vec{\hat \psi};\vec y)) \nabla g^{-1}(\vec{\hat\eta}_{q})J_{N_{q}}\nonumber
\end{align}
where the expression for the matrix $C(\vec \psi;\vec y)=Var(\vec{\hat\eta}_{q})$ is given in the Appendix Section \eqref{predmean} and function $g^{-1}$ is applied to the vector $\vec{\hat\eta}_{q}$ elementwise.
\end{lem}
The proof of Lemma \eqref{var_pred} is given in the Appendix Section \eqref{predmean}. Now, with the variance of the predictor having been approximately estimated, we propose an approximate large sample prediction interval 
\begin{equation}\label{dir.pr}
\hat\lambda_{q}\pm z_{1-\alpha/2} \sqrt{Var (\hat\lambda_{q})}.
\end{equation}
In the future, we will call this one a direct interval. For comparison purposes, we also obtain another prediction interval that will be called, for conciseness, an inverse prediction interval. This interval uses the same predicted group mean \eqref{gr.mean} and its estimated variance \eqref{pr.var}. However, now we will work on the linear predictor scale. To obtain the needed interval, we need, as a first step, to obtain the variance of the link function of the group mean (logit in the logistic case, log in the negative binomial case) using the delta method.  Next, as a second step, the final prediction interval will have the endpoints \begin{equation}\label{pred_inverse_logistic}
(L(\hat\lambda_{q})\,,U(\hat\lambda_{q}))
\end{equation}
similarly to \eqref{lbound} and \eqref{ubound} in the logistic case. In the negative binomial case, the interval will be \begin{equation}\label{pred_inverse_negbin}
(LNB(\hat\lambda_{q}),UNB(\hat\lambda_{q}))
\end{equation}
similarly to \eqref{inv.negb}.

\section{Simulation studies}\label{sim}
	In this section, we illustrate performance of the proposed intervals using synthetic models. We consider both logistic and negative binomial models. 
	For each simulation setting, $5000$ random samples are generated.  
	
\subsection{Logistic Data with One Random Effect}	
	In this section, we consider a logistic regression model with a single Gaussian random effect. This model is commonly used in diabetes clinical trials when the proportion of patients reaching the given treatment target for hemoglobin A1C (HbA$_{1c}$) is the object of interest. First, for $i$th subject, we define $X_{i}$ to be the baseline covariate that is modeled as either a Bernoulli distributed random variable with mean $p=0.5$ or as an independent uniform random variable on $[0,1].$ The Bernoulli baseline covariate can be viewed as some important prognostic factor e.g. whether to use sulfonylurea as a background therapy. The uniform random variable can be viewed as an e.g. the baseline proportion of patients with hypoglycemia exceeding $1.$ Next, $U_{i}$ is the treatment indicator with $0$ meaning control and $1$ meaning experimental treatment. We consider two possible designs of the study. The first design has only two distinct groups (treatment arms): placebo and experimental treatment. Both of these groups are observed at post-randomization time points $1$ and $2.$ In this case, the random effect is used to simulate the within-subject correlation for repeated measurement at different time points. The second study design assumes four distinct groups that are defined by the treatment (experimental or placebo) and subject gender (male or female). In that case, the random effect is used to reflect the fact that experimental subjects are selected at random from a larger population of available subjects. In the definition of the model, the variable $t$ stands for either time or gender. Thus, $t=0$ for the first post-randomization time point, and $t=1$ for the second post-randomization time point. At the same time, we also assume that $t=0$ implies female, and $t=1$ implies male. Also, $\xi\sim N(0,\sigma^{2})$ a Gaussian random effect with $\sigma=0.5.$ We consider an unbalanced design case where the size of each treatment arm is not the same. When observations are recorded over time, this usually happens due to the presence of missing data because of subject drop-out. In practice, the drop-out may occur at any time point; however, subjects without any post-randomization measurement are routinely excluded from clinical trial analyses. Therefore, we assume that missing values can only occur at the second post-randomization time point. In particular, we assume that the sizes are $200$ subjects for experimental treatment group at time zero, $180$ subjects for experimental treatment group at the time point $1,$ again $200$ for control group at the post-randomization time point $0,$ and $160$ for the control group at the postrandomization time point $1.$ This assumption is used to keep missingness rate at $10\%$ for the experimental treatment group and at $20\%$ for the control group. In the case where there are four different treatment arms, the unbalanced design corresponds to the situation where the number of male subjects is lower than the number of female subjects for both control and treatment groups.

	Thus, for $i$th subject the observed data are $Y_{it}$ where 
	\[
	Y_{it}\sim Bern(\mu_{it})
	\] 
	and
	\[
	\log \frac{\mu_{it}}{1-\mu_{it}}=-0.3 -3.0 X_{i}+2U_{i}+0.2t+\xi_{i}.
	\]
   Our chosen  parameter values are almost identical to those chosen by \cite{qu2015estimation} (except the random effect variance since there is no random effect in \cite{qu2015estimation}). The only exception is that our intercept is equal to $-0.3$ instead of $-0.2$.
   The results of our modeling are given in the Table \eqref{table:ci_ul_balanced}. In this Table, the variable $T_{1}$ is an indicator of the type of the baseline covariate used: $T_{1}=1$ corresponds to the use of the Bernoulli random variable with the mean $0.5$ while $T_{1}=2$ corresponds to the use of the $Unif[0,1]$ baseline variable. $T_{2}=1$ corresponds to using the gender as an additional covariate while $T_{2}=2$ means that we are using time covariate. The treatment indicator that comes next, takes on values of either $0$ or $1,$ respectively. The next variable $t$ stands for either time point or the gender, depending on the design of the study. $\mu$ stands for the true population mean for each group computed as in \eqref{true_gr_mean} and \eqref{log.norm}; $\overline{Y}$, observed group mean for patients; $\hat{\mu}$, group mean estimator based on the proposed method and computed according to \eqref{gr_mean}; $\hat{\mu}^{*}$, group mean at the mean baseline computed as 
   \begin{equation}\label{mcov1}
   \hat\mu^{*}=\bE\,g^{-1}(\bar{X}^{'}\hat\beta+b_{i})
   \end{equation} 
   where $\bar X$ is the average of all $\vec x_{i},$ $g$ is the link function (logit in the logistic case),and the expectation is with respect to the distribution of the random effect $b_{i}$ (note that thus defined $\hat{\mu}^{*}$ is conceptually similar to the one defined for the model without a random effect in formula (2) in \cite{qu2015estimation}); CP is the observed coverage probability of each of the two types of the $95\%$ confidence interval used and SD stands for standard deviation. The two types of confidence intervals are CP1 - the inverse interval; CP2 - the direct interval.

The sample mean serves as a benchmark, together with the $\hat{\mu}^{*},$ for our proposed estimator of the group mean. First, note that the performance of the estimator $\hat{\mu}^{*}$ where the mean was estimated at mean covariate level, is highly inadequate. The reason for that is that $\hat{\mu}^{*}$ is not a consistent estimator of $\mu.$ More specifically, the bias $\hat{\mu}^{*}-\mu$ is considerably worse than the bias of our proposed estimator $\hat{\mu}-\mu.$ Both the bias of $\hat\mu$ and the bias of $\bar Y$ are very small in absolute value and so it is important to compare their standard deviations. Here our proposed estimator performs consistently better than the sample mean $\bar Y,$ exhibiting smaller standard deviation for both treatments and time points (gender values). The inverse confidence interval offers a better empirical coverage than the direct interval; as a matter of fact, its empirical coverage seems to be close to the nominal $95\%.$ Both intervals seem to be somewhat downward biased.

The slight downward bias in the empirical probability coverage for both confidence intervals in this case may be due to the underestimation of the estimated group mean variance. This underestimation occurs because the variance of individual $\hat\mu_{ij}$ is found using the approximation \eqref{zeger} for the true $\mu_{ij}.$ It is our conjecture that this slight downward bias can possibly be corrected by using higher-order approximations of the logistic-normal integral  \eqref{log.norm}.

Now, we consider a prediction interval for the group mean $\lambda$ defined as in \eqref{gr.mean}. We also use the notation $\hat\lambda$ for the predicted group mean computed as in \eqref{pred.gr.mean}. Finally, let the point predictor at the mean baseline for $\eta_{i}$ be defined as $\eta_{i}^{*}(\vec\psi;\vec y):={\bar x}^{'}\vec\beta+\vec z_{i}^{'}\vec b(\vec\psi;\vec y).$ Then, the predictor at the mean covariate is 
\begin{equation}\label{mcov2}
\hat\lambda^{*}=\frac{1}{N_{q}}\sum_{i\in q}g^{-1}(\eta_{i}^{*}(\vec{\hat\psi};\vec y)).
\end{equation} We consider two types of prediction intervals. The first one, denoted CP1, is the inverse prediction interval that is constructed according to \eqref{pred_inverse_logistic}. The second one, denoted CP2, is the direct confidence interval that is constructed according to \eqref{dir.pr}. The outcome of this simulation study is summarized in the Table \eqref{table:pi_ul_balanced}.		
	
Note that the empirical coverage probability of the inverse prediction interval does not seem to be downward biased as opposed to confidence intervals using Zeger's approximation for the logistic-normal integral. In fact, it is close to the nominal coverage probability of $95\%.$ The empirical coverage probability of the direct interval is slightly worse and may be exhibiting some downward bias. Although the bias of the predicted group mean $\hat\lambda$ has not decreased in comparison with that of the raw estimator defined by the sample mean $\bar Y,$ the standard deviation of the predicted group mean $\hat\lambda$ is consistently smaller.

	\subsection{Negative Binomial Data with One Random Effect}	
	
	In this section, we consider a negative binomial model. Such a model has been used rather extensively to describe the number of hypoglycemic events in diabetes clinical trials since the distribution for the number of hypoglycemic events is known to be generally skewed with the variance exceeding the mean (see e.g. \cite{qu2015estimation}). Since the negative binomial distribution can be thought of as an overdispersed Poisson distribution \cite{mccullagh2018generalized}, it seems to be a very appropriate choice for modeling count data with variance exceeding the mean. The notation is almost the same as the one used earlier for the logistic model. In particular, for $i$th subject, we define $X_{i}$ to be the baseline covariate that is modeled as either a Bernoulli distributed random variable with mean $p=0.5$ or as an independent uniform random variable on $[0,1].$ Once again, $U_{i}$ is the treatment indicator. The first study design considered has only two distinct groups (treatment arms): placebo and experimental treatment. Both of these groups are observed at post-randomization time points $0$ and $1$. The second study design assumes four distinct groups that are defined by the treatment (experimental or placebo) and subject gender (male or female). In the definition of the model, the variable $t$ stands for either time or gender with the same possible values as in the case of logistic model. Finally, $\xi\sim N(0,\sigma^{2})$ a Gaussian random effect with $\sigma=0.1.$ In this study, the generated data are $Y_{it}$ that have the negative binomial distribution and are assumed to model the number of hypoglycemic events for $i$th subject with the indicator $t$. Thus, $Y_{it}\sim NB(\mu_{it},\kappa)$ where $\bE (Y_{it})=\mu_{it}$ and $Var (Y_{it})=\mu_{it}+\frac{\mu_{it}^{2}}{\kappa}$ with $\kappa$ being an unknown parameter that is commonly called the {\it size} of the negative binomial distribution. The mean is modeled as 
	\[
	\log\left(\mu_{it}\right)=0.3-0.2X_{i}+0.3U_{i}+0.4t+\xi_{i}.
		\]	
The size parameter is assumed to be $\kappa=50.$ We make the same assumptions of $200$ and $180$ subjects for the experimental treatment group at times $0$ and $1$, respectively, and $200$ and $160$ subjects for the placebo group at times $0$ an $1,$ respectively. In case where there are four different treatment arms, this design corresponds to the situation where the number of male subjects is lower than the number of female subjects for both control and treatment groups.The results of our modeling are given below in the Table \eqref{table:ci_nb_unbalanced}. This table provides empirical coverage for three different types of confidence intervals: CP1 is the inverse confidence interval as defined in \eqref{inv.negb}, CP2 is the direct confidence interval as defined in \eqref{CI_negb1}, and CP3 is the lognormal confidence interval as defined in \eqref{lognormal.negb}. Looking at the Table \eqref{table:ci_nb_unbalanced}, we note that the coverage probabilities of all the confidence intervals seem to be better than those for the two intervals available in the logistic case; this is, most likely, the result of having to use an approximation for the logistic normal integral when constructing confidence intervals in the logistic case. None of the intervals show much evidence of any downward or upward bias in their coverage probabilities. All of the intervals considered seem to demonstrate coverage probability that is close to the nominal $95\%$ for all groups. The proposed estimator $\hat{\mu}$ displays rather small bias for nearly all considered settings, that is across both treatment groups and time points/gender covariate values. The bias of $\hat\mu$ is similar in magnitude to the bias of the raw estimator $\bar Y;$ however, the standard deviation of $\hat\mu$ is consistently smaller. It performs considerably better than the benchmark estimator $\hat{\mu}^{*}$ of \cite{qu2015estimation} by taking into account the presence of a random effect; this can be seen clearly by comparing the bias of the two estimators. The estimator  $\hat{\mu}^{*}$ displays substantial bias because it is not consistent as an estimator of $\mu.$

Finally, we also construct prediction intervals for the conditional group mean $\lambda$ in the negative binomial setting as well. In this case, we can only consider two possible prediction intervals for the predicted conditional group mean $\hat\lambda$ - the inverse one, denoted by CP1, and the direct one, denoted by CP2. Simulation results are presented in the Table \eqref{table:pi_nb_unbalanced}. The proposed prediction $\hat\lambda$ has consistently smaller standard deviation than the raw estimator $\bar Y.$ Moreover, the bias of $\hat\lambda$ is consistently smaller than the bias of the estimator at mean covariate $\hat\lambda^{*}.$ However, in both of these cases, the empirical coverage probability of both prediction intervals (especially the direct one) appears somewhat downward biased. 
	
\section{Real Data}	

Both confidence interval and prediction interval based methods were applied to a 24-week, multicenter, open-label diabetes clinical trial for patients with type 2 diabetes mellitus. This dataset originates in \cite{rosenstock2008advancing} and has also been described and used in \cite{qu2015estimation}; thus, we will be brief in its description. In total, $374$ patients took basal insulin glargine at baseline and were, then, randomly assigned to take either lispro mix 50/50 (LM) or basal bolus therapy (BBT). The purpose was the comparison of hypoglycemia between the two treatment groups for the titration period (first $12$ weeks) and the maintenance period (the last 12 weeks). The four variables considered were the total and nocturnal hypoglycemia rates per $30$ days as well as the total and nocturnal hypoglycemia incidence. The hypoglycemia incidence was analyzed using a logistic GLMM with a single subject-wise random effect. Hypoglycemia rates were analyzed using a negative binomial GLMM with a single subject-wise random effect. Again, the baseline hypoglycemia rate was used as a baseline variable. 

The result of our analyses is shown in Table \eqref{table:real_glmm}. We obtain both unconditional estimated mean $\hat\mu$ computed according to the formula \eqref{gr_mean}, and the conditional estimated mean $\hat\lambda,$ computed according to formula \eqref{gr.mean}, for each group. For each group, both of these means are given together with their standard errors. We also report the raw mean $\bar Y$ for each group as well as the mean estimated at mean covariate $\hat\mu^{*}.$  We note that, in the same way as in \cite{qu2015estimation}, both conditional and unconditional estimated group means are consistently larger than the mean estimated at mean covariate $\hat\mu^{*}$ for both total and nocturnal hypoglycemia rate modelled using negative binomial GLMM. The difference is much smaller for logistic models that are used for modeling proportion of patients with either total or nocturnal hypoglycemia rate. This suggests that the estimate for the mean at mean covariate $\hat\mu^{*}$ can, in practice, be considerably smaller than both conditional and unconditional means suggested by our method. The conditional estimated group mean is consistently larger than the unconditional one for negative binomial models; the two means are, however, very close to each other (and to the mean at mean covariate) when it comes to logistic model.

\section{Summary and Discussion}\label{sum}

Presence of subject-wise random effect is rather common in medical data. This is especially true when it comes to data of longitudinal nature where the presence of such an effect can be thought to follow directly from the dependence of the data points over time. In other situations, such a random effect may also be necessary to model a randomized selection of the trial subjects from a large pool of candidates. In a situation where such a random effect has to be accounted for, we proposed two possible estimators of group means for the study population in clinical trials. The first of these estimators is unconditional and is based on the averaging the random effect over its distribution. The other one is conditional on the data and is based on the predicted value of the random effect.  For both proposed estimators, we obtained approximate variances and constructed $95\%$ confidence and prediction intervals, respectively. 
Most of the confidence and prediction intervals we constructed show excellent empirical coverage probability that is close to $95\%.$ Thus, we believe that both approaches we suggested should be useful when estimating/predicting the group mean for the study population in clinical trials. One example of a slight (downward) bias in the empirical coverage probability is given by confidence intervals for the group means in the logistic model. We note that this is most likely caused by the use of a relatively simple approximation of the logistic-normal integral. This suggests, as a practical matter, that for the logistic case, the use of prediction intervals should be preferred for now. One of our future research directions will involve correction of the bias in empirical coverage probabilities for confidence intervals in logistic models through the use of higher-order approximations of the logistic-normal integral. The entire estimation and prediction procedure used in this manuscript has been implemented in R, using to a great extent a very convenient package lme4. Many elements of the computational procedure, such as, for example, calculation of conditional predictions and their variances, have been programmed in the form of convenient routines. The authors plan  to combine most of the resulting procedures into an easy to use R package that will be disseminated online for general use in the near future.



\section{Supplementary materials}

The reader is referred to the on-line Supplementary Materials for technical appendices and annotated R programs. The data that support the findings of this study are available from Eli Lilly. Restrictions apply to the availability of these data, which were used under license for this study. All of our codes are available at \url{https://github.com/duanjiexin/GLMM_GM}

\section{Acknowledgements}

The authors would like to express their gratitude to Eli Lilly for its financial support of our research through the grant 40002375. We also want to express our thanks to Prof. James Booth of the Department of Statistical Science at Cornell University for several useful discussions.

\bibliographystyle{asa}
\bibliography{Lilly}

\begin{thebibliography}{21}
\newcommand{\enquote}[1]{``#1''}
\expandafter\ifx\csname natexlab\endcsname\relax\def\natexlab#1{#1}\fi

\bibitem[{Bartlett(2018)}]{bartlett2018covariate}
Bartlett, J.~W. (2018), \enquote{Covariate adjustment and estimation of mean
  response in randomised trials,} \textit{Pharmaceutical Statistics}, 17,
  648--666.

\bibitem[{Bibby et~al.(1979)Bibby, Kent, and Mardia}]{bibby1979multivariate}
Bibby, J., Kent, J., and Mardia, K. (1979), \enquote{Multivariate analysis,} .

\bibitem[{Booth and Hobert(1998)}]{booth1998standard}
Booth, J.~G. and Hobert, J.~P. (1998), \enquote{Standard errors of prediction
  in generalized linear mixed models,} \textit{Journal of the American
  Statistical Association}, 93, 262--272.

\bibitem[{De~Bruijn(1981)}]{de1981asymptotic}
De~Bruijn, N.~G. (1981), \textit{Asymptotic methods in analysis}, vol.~4,
  Courier Corporation.

\bibitem[{Demidenko(2013)}]{demidenko2013mixed}
Demidenko, E. (2013), \textit{Mixed models: theory and applications with R},
  John Wiley \& Sons.

\bibitem[{Jiang(2007)}]{jiang2007linear}
Jiang, J. (2007), \textit{Linear and generalized linear mixed models and their
  applications}, Springer Science \& Business Media.

\bibitem[{Jorgensen(1987)}]{jorgensen1987exponential}
Jorgensen, B. (1987), \enquote{Exponential dispersion models,} \textit{Journal
  of the Royal Statistical Society. Series B (Methodological)}, 49, 127--162.

\bibitem[{Kotz et~al.(2000)Kotz, Balakrishnan, and Johnson}]{kotzcontinuous}
Kotz, N., Balakrishnan, N., and Johnson, N. (2000), \enquote{Continuous
  Multivariate Distributions, Volume 1: Models and Applications,} .

\bibitem[{Lalonde and Qu(2019)}]{LalQu}
Lalonde, A. and Qu, Y. (2019), \enquote{Estimation of Group Means Using
  Bayesian Generalized Linear Mixed Models,} Submitted.

\bibitem[{McCullagh(2018)}]{mccullagh2018generalized}
McCullagh, P. (2018), \textit{Generalized linear models}, Routledge.

\bibitem[{Pinheiro and Bates(1995)}]{pinheiro1995approximations}
Pinheiro, J.~C. and Bates, D.~M. (1995), \enquote{Approximations to the
  log-likelihood function in the nonlinear mixed-effects model,}
  \textit{Journal of Computational and Graphical Statistics}, 4, 12--35.

\bibitem[{Pratesi et~al.(2000)Pratesi, Santucci, Graziosi, and
  Ruggieri}]{pratesi2000outage}
Pratesi, M., Santucci, F., Graziosi, F., and Ruggieri, M. (2000),
  \enquote{Outage analysis in mobile radio systems with generically correlated
  log-normal interferers,} \textit{IEEE Transactions on communications}, 48,
  381--385.

\bibitem[{Qu and Luo(2015)}]{qu2015estimation}
Qu, Y. and Luo, J. (2015), \enquote{Estimation of group means when adjusting
  for covariates in generalized linear models,} \textit{Pharmaceutical
  Statistics}, 14, 56--62.

\bibitem[{Rio(1996)}]{rio1996theoreme}
Rio, E. (1996), \enquote{Sur le th{\'e}oreme de Berry-Esseen pour les suites
  faiblement d{\'e}pendantes,} \textit{Probability theory and related fields},
  104, 255--282.

\bibitem[{Rosenstock et~al.(2008)Rosenstock, Ahmann, Colon, Scism-Bacon, Jiang,
  and Martin}]{rosenstock2008advancing}
Rosenstock, J., Ahmann, A.~J., Colon, G., Scism-Bacon, J., Jiang, H., and
  Martin, S. (2008), \enquote{Advancing insulin therapy in type 2 diabetes
  previously treated with glargine plus oral agents: prandial premixed (insulin
  lispro protamine suspension/lispro) versus basal/bolus (glargine/lispro)
  therapy,} \textit{Diabetes care}, 31, 20--25.

\bibitem[{Safak and Safak(1994)}]{safak1994moments}
Safak, A. and Safak, M. (1994), \enquote{Moments of the sum of correlated
  log-normal random variables,} in \textit{Vehicular Technology Conference,
  1994 IEEE 44th}, IEEE, pp. 140--144.

\bibitem[{Schall(1991)}]{schall1991estimation}
Schall, R. (1991), \enquote{Estimation in generalized linear models with random
  effects,} \textit{Biometrika}, 78, 719--727.

\bibitem[{Schwartz and Yeh(1982)}]{schwartz1982distribution}
Schwartz, S.~C. and Yeh, Y.-S. (1982), \enquote{On the distribution function
  and moments of power sums with log-normal components,} \textit{Bell System
  Technical Journal}, 61, 1441--1462.

\bibitem[{StataCorp(2013)}]{statacorp2013stata}
StataCorp, L. (2013), \enquote{Stata multilevel mixed-effects reference
  manual,} College Station, TX: StataCorp LP.

\bibitem[{Stroup(2016)}]{stroup2016generalized}
Stroup, W.~W. (2016), \textit{Generalized linear mixed models: modern concepts,
  methods and applications}, CRC press.

\bibitem[{Zeger et~al.(1988)Zeger, Liang, and Albert}]{zeger1988models}
Zeger, S.~L., Liang, K.-Y., and Albert, P.~S. (1988), \enquote{Models for
  longitudinal data: a generalized estimating equation approach,}
  \textit{Biometrics}, 44, 1049--1060.

\end{thebibliography}

\renewcommand{\theequation}{S.\arabic{equation}}
\renewcommand{\thetable}{S\arabic{table}}
\renewcommand{\thefigure}{S\arabic{figure}}
\renewcommand{\thesection}{S.\arabic{section}}
\renewcommand{\thesubsection}{S.\arabic{section}.\arabic{subsection}}
\setcounter{equation}{0}
\setcounter{table}{0}
\setcounter{section}{0}
\setcounter{subsection}{0}

\newtheorem{theoremA}{Theorem}
\renewcommand{\thetheoremA}{A.\arabic{theoremA}}
\newtheorem{lemmaS}{Lemma}
\renewcommand{\thelemmaS}{S.\arabic{lemmaS}}



\newpage

\begin{center}
\Large\bf Supplement to: Estimation of Group Means in Generalized Linear Mixed Models
\end{center}

{
\begin{center}
    Jiexin Duan, Michael Levine, Junxiang Luo and Yongming Qu
\end{center}
}



\section{Appendix}

\subsection{General form of the marginal likelihood for GLMMs based on exponential families}\label{marglik}

In general, when estimating parameters $\vec\beta$ and $\sigma^{2},$ it is useful to keep in mind that the logistic, negative binomial, Poisson, and many other generalized linear mixed models, can be derived from a two-parameter canonical exponential family with the density
\begin{equation}\label{expf}
f(y;\theta,\phi)=\exp\left[\frac{\theta y-v(\theta)}{\phi}-t(y,\phi)\right]
\end{equation} where $\theta$ and $\phi$ are parameters. $\theta$ is typically called the linear predictor and $\phi$ the scale parameter. Such a model implies that the mean is equal to $v^{'}(\theta)$, and the variance to $v^{''}(\theta)$. In most cases, the variance is a simple function of the mean. For example, for binary data we have the variance equal to $\mu(1-\mu)$, the scale parameter $\phi=1$ and the distribution in \eqref{expf} simplifies to $f(y;\theta)=e^{\theta y-v(\theta)-t(y)}$.  In order to avoid confusion when integrating out the random effect, we denote a generic subject-wise random effect $u.$ The marginal likelihood (where the random effect is integrated out) based on all $N$ observations is, then, 
\begin{equation}\label{mrgl}
l(\vec \beta,\sigma^{2})=-\frac{N}{2}\log(2\pi)-\frac{N}{2}\log\sigma^{2}+\sum_{i=1}^{N}\log \int e^{l_{i}(\vec \beta,u)-\frac{u^{2}}{2\sigma^{2}}\,du}
\end{equation}
where 
\[
l_{i}(\vec \beta,u)=\sum_{j=1}^{N_{i}}[(\vec x_{ij}^{'}\vec \beta+u)y_{ij}-v(\vec x_{ij}^{'}\vec\beta+u)]
\]
is the $i$th conditional log-likelihood.

\subsection{Estimation of the variances and covariances of individual subject marginal means $\hat\mu_{i}$}\label{margmean}
To begin with, we will find the variance of $\hat \mu_{i}$ using the multivariate delta method. Note that using Zeger's approximation allows us to compute the approximate gradient of $\hat\mu_{i}$ explicitly. Using the notation $\vec \psi=(\vec \beta^{'},\sigma)^{'}$, and corresponding $\hat{\vec \psi}=(\hat{\vec \beta}^{'},\hat\sigma)^{'}$, we find that the gradient of $\hat \mu_{i}$ with respect to $\hat{\vec \beta}$ and $\hat\sigma$ is $\nabla_{\hat{\vec \psi}} \hat \mu_{i}=\left(\frac{\partial \hat{\mu}_{i}}{\partial \hat{\vec\beta}}\,, \frac{\partial \hat{\mu}_{i}}{\partial  \hat\sigma}\right)^{'}$, where
\[
\frac{\partial \hat{\mu}_{i}}{\partial \hat{\vec \beta}} \approx \frac{\exp(\hat{c}\vec{x}_{i}^{'} \hat{\vec \beta})}{\left[ 1+\exp(\hat{c}\vec{x}_{i}^{'}\hat{\vec \beta})\right]^2} \hat{c}\vec{x}_{i} = \frac{\exp(\hat{c}\vec{x}_{i}^{'} \hat{\vec \beta}) \hat{c}\vec{x}_{i}}{\left[ 1+\exp(\hat{c}\vec{x}_{i}^{'}\hat{\vec \beta})\right]^2} 
\]
\[
\frac{\partial \hat{\mu}_{i}}{\partial \hat{\sigma}} \approx \frac{-0.346\hat {\sigma} \hat c^{3} \exp(\hat{c}\vec{x}_{i}^{'} \hat{\vec \beta})  \vec{x}_{i}^{'} \hat{\vec \beta}}{\left[ 1+\exp(\hat{c}\vec{x}_{i}^{'}\hat{\vec \beta})\right]^2}
\]
Thus, the variance of $\hat{\mu}_{i}$ is approximately equal to 
\[
Var\,(\hat \mu_{i})\approx \nabla^{'}_{\hat{\vec \psi}} \hat \mu_{i}\Sigma_{\hat{\vec \psi}} \nabla_{\hat{\vec \psi}} \hat \mu_{i}
\]
where $\Sigma_{\hat{\vec \psi}}$ is the covariance matrix of the estimated parameter vector $\hat \psi$. We will use the empirical Fisher scoring algorithm to compute maximum likelihood estimates of $\hat{\vec \beta}$ and $\sigma^{2}$; see \cite{demidenko2013mixed}, pp. $363-364$ for details. In brief, this algorithm uses the sum of squared score functions $\vec d_{i}$ for the $i$th subject, $i=1,\ldots,K,$ to estimate the information matrix at each iteration; that is, $H=\sum_{i=1}^{K}\vec d_{i}\vec d_{i}^{'}$. Therefore, the inverse Hessian matrix $H^{-1}$, evaluated at the final iteration, provides a consistent estimate of the covariance matrix $\Sigma_{\hat \psi}$. 

As a second step, we now estimate the covariance $Cov \left(\hat \mu_{i_{1}}, \hat \mu_{i_{2}}\right)$. This can be done using the multivariate delta method again. More specifically, we approximate the covariance within $q$th group as 

\begin{equation}\label{cov1}
Cov \left(\hat \mu_{i_{1}}, \hat \mu_{i_{2}}\right)\approx \nabla^{'}_{\hat{\vec \psi}} \hat \mu_{i_{1}}\Sigma_{\hat{\vec \psi}}\nabla_{\hat{\vec \psi}} \hat \mu_{i_{2}}
\end{equation}
for any $i_{1}<i_{2}$ such that $i_{1},i_{2}\in q.$

\subsection{Approximation of the marginal group mean in the negative binomial case}\label{negbin}

To obtain the approximation \eqref{varmu}, we utilize the following result from \cite{kotzcontinuous}.
\begin{lemmaS}\label{kotz}
Let $\Lambda=\sum_{i=1}^{n}\exp (X_{i})$ where $\vec X=\{X_{i}\}_{i=1}^{n}$ is a multivariate normal distribution with the mean vector $\vec \mu=(\mu_{1},\ldots,\mu_{n})^{'}$ and the covariance matrix $M=\{M_{i,j}\}_{i,j=1}^{n}$ where $M_{i,j}=Cov (X_{i},X_{j})=\sigma^{2}_{ij}$. Then, the variance $Var \Lambda=\sum_{i,j=1}^{n}Cov (\exp (X_{i}),\exp(X_{j}))=\sum_{i,j=1}^{n}\exp(\mu_{i}+\mu_{j}+\frac{1}{2}(\sigma_{ii}^{2}+\sigma_{jj}^{2}))(\exp (\sigma^{2}_{ij})-1)$. 
\end{lemmaS}
The Lemma \eqref{kotz} will enable us to compute an approximate variance of $N_{q}\hat\mu_{q}=\sum_{i\in q}\hat \mu_{i}$.  Recall that $\hat \nu_{i}=\log \hat\mu_{i}=\vec x_{i}^{'}\hat\beta+\hat\sigma^{2}/2$. The asymptotic distribution of $\hat\sigma^{2}$ can be viewed as approximately $\chi^{2}$ under the so-called ``small-dispersion" assumption as defined in e.g. \cite{jorgensen1987exponential}. Roughly speaking, ``small-dispersion" asymptotics implies that the total sample size $N\rightarrow \infty$ and the smallest group size $\min_{i\in q}(N_{q})\rightarrow \infty$ as well. The $\chi^{2}$ distribution involved can be viewed as approximately normal for a large number of degrees of freedom that will be the case in a typical study with a large number of subjects. Moreover, $\hat{\vec{\beta}}$ will have an asymptotically normal distribution; for an example of such an asymptotic result see e.g. \cite{demidenko2013mixed} p. $381.$ Therefore, each $\hat\nu_{i}$ can be viewed as approximately normally distributed. Moreover, the same asymptotic result suggests that the estimator of $\hat\nu_{i}$ is asymptotically unbiased and so the mean of the  normal distribution of $\hat\nu_{i}$ is, approximately, $\nu_{i}=\vec{x}_{i}^{'}\vec \beta+\sigma^{2}/2$. Using the argument above, we can view each estimator $\hat\mu_{i}$ as approximately lognormally distributed. To construct a confidence interval, we will also need to estimate each element of the covariance matrix $M=Cov(\hat\nu_{i_{1}},\hat\nu_{i_{2}})$ with $i_{1}<i_{2}$ such that $i_{1},i_{2}\in q.$ This can be done by using the delta method. The gradient $\nabla_{\hat\psi}\hat\nu_{i}=(\vec x_{i}^{'},\hat\sigma)^{'}$ and e.g. the approximate variance $\hat\sigma^{2}_{i}:=Var\,\hat\nu_{i}\approx \nabla_{\hat\psi}^{'}\hat\nu_{i}\Sigma_{\hat\psi}\nabla_{\hat\psi}\hat\nu_{i}$ while the approximate covariance of any two $\hat\nu_{i_{1}},$ $\hat\nu_{i_{2}}$ within $q$th group is $\hat\sigma^{2}_{i_{1},i_{2}}:=Cov\,(\hat\nu_{i_{1}},\hat\nu_{i_{2}})\approx \nabla_{\hat\psi}^{'}\hat\nu_{i_{1}}\Sigma_{\hat\psi}\nabla_{\hat\psi}\hat\nu_{i_{2}}.$ We can now approximate the variance of $\hat\mu_{q}$ by using the lemma \eqref{kotz} and putting together the above approximations.  

\subsection{Approximate variance of the predicted group mean}\label{predmean}
If the parameter vector $\vec\psi$ is known, a sensible prediction variance for $\vec\eta_{i}$ is 
\begin{equation}\label{pred.var}
\nu_{i}(\vec \psi;\vec y):=\vec z_{i}^{'}Var_{\vec\psi}(\vec b|\vec y)\vec z_{i}.
\end{equation}
Of course, in practice $\vec\psi$ is not known and we have to plug in its stead some consistent estimator $\vec{\hat\psi}.$ Simply substituting $\vec{\hat\psi}$ into \eqref{pred.var} does not work since this approach ignores sampling variability associated with $\vec{\hat\psi}.$  Following the idea of \cite{booth1998standard}, we define the prediction variance of $\hat\eta_{i}=\eta_{i}(\vec{\hat \psi};\vec y$) as the conditional squared error $\hat\eta_{i}$:

\begin{equation}\label{cond.var}
C_{i}(\vec\psi;\vec y)=\bE_{\vec\psi}\{(\hat\eta_{i}-\eta_{i})^{2}|\vec y\}.
\end{equation}
Since $\eta_{i}(\vec\psi;\vec y)-\eta_{i}$ and $\hat \eta_{i}-\eta_{i}(\vec\psi;\vec y)$ are independent, one can easily show that 
\begin{align*}
&C_{i}(\vec\psi;\vec y)=\nu_{i}(\vec \psi;\vec y)+\bE_{\vec\psi}\{\{\hat\eta_{i}-\eta_{i}(\vec\psi;\vec y)\}^{2}|\vec y\}\\
&=\nu_{i}(\vec \psi;\vec y)+c_{i}(\vec\psi;\vec y)
\end{align*}
where $c_{i}(\vec\psi;\vec y)$ is a nonnegative correction term that accounts for estimation of $\vec\psi$ while the term $\nu_{i}(\vec \psi;\vec y)$ is the so-called ``naive" prediction variance of $\hat\eta_{i}$ that would have been used if $\vec\psi$ was known. Using Laplace approximation as in, for example, \cite{de1981asymptotic}, chap. $4$ (see also \cite{booth1998standard}) one can obtain the following approximation to the ``naive" prediction variance of $\hat\eta_{i}:$
\[
\nu_{i}(\vec\psi;\vec y)\approx \vec z_{i}^{'}(Z^{'}\tilde W Z+{\cal G}^{-1})^{-1}\vec z_{i}. 
\]
Now we need to provide a convenient approximation to the second term $c_{i}(\vec\psi;\vec y)$ that accounts for the sampling variability due to parameter estimates. As a first step, we note that 
\begin{align*}
&\hat\eta_{i}-\eta_{i}(\vec\psi;\vec y)=\vec x_{i}^{'}(\vec{\hat\beta}-\vec\beta)+\vec z_{i}^{'}(\vec b(\vec{\hat\psi};\vec y)-\vec b(\vec\psi;\vec y))\\
&=\vec x_{i}^{'}(\vec{\hat\beta}-\vec\beta)+\vec z_{i}^{'}\left\{\frac{\partial \vec b(\vec \psi;\vec y)}{\partial \vec\psi}\left(\vec{\hat\psi}-\vec\psi\right)+O_{p}\left(|\vec{\hat\psi}-\vec\psi|^{2}\right)\right\}
 \end{align*}
Since $\vec\psi=(\vec\beta^{'},\vec{(\sigma^{2})}^{'}),$ we can write the $K\times (p+2)$ matrix $\frac{\partial \vec b(\vec \psi;\vec y)}{\partial \vec\psi}$ as a combination of two block matrices
\[
\frac{\partial \vec b(\vec \psi;\vec y)}{\partial \vec\psi}=\left(\frac{\partial \vec b(\vec \psi;\vec y)}{\partial \vec\beta},\frac{\partial \vec b(\vec \psi;\vec y)}{\partial \vec\sigma^{2}}\right).
\]
The above implies that 
\[
\hat\eta_{i}-\eta_{i}(\vec\psi;\vec y)\approx A_{i}(\vec\psi;\vec y)^{'}(\vec{\hat\psi}-\vec\psi)
\]
 where 
 \begin{equation}\label{aform}
 A_{i}(\vec\psi;\vec y)=
 \begin{bmatrix}
           \vec x_{i}+\left( \frac{\partial \vec b(\vec \psi;\vec y)}{\partial \vec\beta} \right)^{'}\vec z_{i} \\
           \left(\frac{\partial \vec b(\vec \psi;\vec y)}{\partial \vec\sigma^{2}}\right)^{'}\vec z_{i}.
                    \end{bmatrix}
\end{equation}

This implies immediately the following approximation for the term $c_{i}(\vec \psi;\vec y):$
\begin{equation}\label{cor_term}
 c_{i}(\vec \psi;\vec y) \approx  A_{i}^{'}(\vec\psi;\vec y)I(\vec\psi)^{-1}A_{i}(\vec\psi;\vec y)
 \end{equation}
 where $I(\vec\psi)$ is the information matrix for the parameter vector $\vec\psi.$ Of course, in practice the expression \eqref{cor_term} has to be estimated by 
 \[
 c_{i}(\vec{\hat\psi};\vec y)= A_{i}^{'}(\vec{\hat\psi};\vec y)I(\vec{\hat\psi})^{-1}A_{i}(\vec{\hat\psi};\vec y);
 \] however, this estimate is rather complicated, and so we will approximate it with an easier to handle expression that is convenient to use for computational purposes. To achieve this, one has to be able  to compute derivatives of the conditional expectation $b_{i}(\vec\psi;\vec y_{i}).$ It can be defined implicitly from the equation
\[
F_{i}(b_{i};\vec\psi)=\frac{\partial}{\partial b_{i}}l_{i}(b_{i};\vec \psi)\approx 0. 
\] 
It follows immediately that 
\[
\frac{\partial b_{i}(\vec\psi;\vec y_{i})}{\partial \vec\psi}\approx -\left(\frac{\partial F_{i}}{\partial b_{i}}\right)^{-1}\frac{\partial F_{i}}{\partial \vec\psi}.
\] The above approximations would have been correct if our model was a linear mixed model (LMM) with normally distributed data; for non-normal data modeled with GLMM, these are approximations only. Let us denote $n_{i}\times 1$ vector of $1$s $J_{i}.$ It is now easy to check directly that $\frac{\partial F_{i}}{\partial b_{i}}=-J_{i}^{'}W_{i}J_{i}-\frac{1}{\sigma^{2}},$ $\frac{\partial F_{i}}{\partial \vec\beta}=-J_{i}^{'}W_{i}X_{i},$ and $\frac{\partial F_{i}}{\partial \sigma^{2}}=\frac{b_{i}}{\sigma^{4}}.$ In the binomial model, the dispersion parameter $\sigma_{0}^{2}\equiv 1$ and therefore, the last remaining derivative $\frac{\partial F_{i}}{\partial \sigma_{0}^{2}}$ is set to zero. All of these derivatives are evaluated at $b_{i}=b_{i}(\vec\psi;\vec y_{i}).$ If the dispersion parameter is not equal to zero (as is the case with negative binomial model), then $\frac{\partial F_{i}}{\partial \sigma_{0}^{2}}=-\frac{1}{\sigma_{0}^{4}}\sum_{j=1}^{n_{i}}w_{ij}(y_{ij}-\lambda_{ij}).$   The above also implies that  
\begin{equation}
\frac{\partial \hat b_{i}}{\partial \vec\beta}=-\left(J_{i}^{'}\tilde W_{i}J_{i}+\frac{1}{\sigma^{2}}\right)^{-1}J_{i}^{'}\tilde W_{i}X_{i}.
\end{equation}
To simplify our approximation of the estimated correction term, we assume that the variance components $\sigma_{0}^{2}$ and $\sigma^{2}$ are known; thus, we do not adjust for the sampling variability of estimated variance components. This assumption is quite common in the GLMM literature; see, for example, Appendix A.1 of \cite{booth1998standard}. In that case, the entire information matrix, $I(\vec\psi),$ becomes $I(\vec\beta),$ which  we need to approximate. 
Using the approach of \cite{booth1998standard} Appendix A.1 and Appendix A.2, the expected information matrix $I(\vec\beta)$ can be approximated as   
\[
I(\vec\beta) \approx X^{'}(\tilde W^{-1}+Z{\cal G}Z^{'})^{-1}X,
\]
and the approximated conditional variance for the linear predictor of the $i$th subject can be written as
\begin{align}\label{cmsep}
&C_{i}(\vec \psi;\vec y)\approx \vec z_{i}^{'}\left(Z^{'}\tilde W Z+{\cal G}^{-1}\right)^{-1}\vec z_{i}+\vec x_{i}^{'}I(\vec \beta)^{-1}\vec x_{i}\\
&-2\vec x_{i}^{'}I(\vec\beta)^{-1}X^{'}\tilde W Z\left(Z^{'}\tilde W Z+{\cal G}^{-1}\right)^{-1}\vec z_{i}\nonumber.
\end{align}

Adding up equations \eqref{cmsep} over all subjects in $q$th group, and using standard matrix identities (see e.g. \cite{bibby1979multivariate}), we will get  
the conditional covariance matrix of the predicted vector $\hat{\vec \eta_{q}}=(\hat\eta_{1},\ldots,\hat\eta_{N_{q}})^{'}$ as   
\[
C(\vec \psi;\vec y) \approx (X_{q}^{'}, Z_{q}^{'})\begin{pmatrix} X^{'}{\tilde W} X & X^{'}{\tilde W} Z\\
Z^{'}{\tilde W} X & Z^{'}{\tilde W} Z+{\cal G}^{-1}\end{pmatrix}^{-1} \begin{pmatrix} X_{q}\\
Z_{q}\end{pmatrix}
\]

It remains finally to note that predicted group mean for the $q$th group is $\frac{1}{N_{q}}\sum_{i=1}^{N_{q}}g^{-1}(\hat\eta_{i})=\frac{1}{N_{q}}J_{N_{q}}^{'}g^{-1}(\hat{\vec \eta_{q}})$ where the function $g^{-1}(\cdot)$ is applied to the vector $\hat{\vec\eta_{q}}$ elementwise.

\section{Tables}

\begin{table*}[!b]	
	\caption{Summary of various marginal group mean estimators for the logistic design case together with relevant confidence intervals. The number of simulations used is $5000.$}
	\label{table:ci_ul_balanced}
	\vspace{-0.5em}
	\begin{center}
		\setlength\tabcolsep{4.0pt} 
		\footnotesize
		\begin{tabular}{cc cc r rr rr rr rr}
			\toprule
			\multicolumn{2}{c}{Type} &\multicolumn{2}{c}{Group}&\multicolumn{1}{c}{$\mu$}&\multicolumn{2}{c}{$\bar{Y}-\mu$}&\multicolumn{2}{c}{$\hat{\mu}^*-\mu$}&\multicolumn{2}{c}{$\hat{\mu}-\mu$}&\multicolumn{2}{c}{$\hat{\mu}$}\\ 
			 $T_1$ & $T_2$ & $U$ & $t$ & Mean & Mean & SD & Mean & SD & Mean & SD & CP1 &CP2 \\ 
			\midrule
 1 & 1 & 1 & 0 & 0.530 &  0.0001 & 0.027 &  0.0185 & 0.039 &  0.0002 & 0.024 & 0.951 & 0.951 \\
 1 & 1 & 1 & 1 & 0.559 &  0.0010 & 0.030 &  0.0359 & 0.041 &  0.0006 & 0.026 & 0.939 & 0.938 \\
 1 & 1 & 0 & 0 & 0.234 & -0.0002 & 0.027 & -0.0837 & 0.023 & -0.0006 & 0.023 & 0.946 & 0.942 \\
 1 & 1 & 0 & 1 & 0.262 & -0.0004 & 0.030 & -0.0847 & 0.027 & -0.0002 & 0.025 & 0.946 & 0.941 \\
 1 & 2 & 1 & 0 & 0.530 &  0.0000 & 0.027 &  0.0180 & 0.040 & -0.0002 & 0.024 & 0.930 & 0.928 \\
 1 & 2 & 1 & 1 & 0.560 &  0.0008 & 0.029 &  0.0353 & 0.040 & -0.0001 & 0.025 & 0.935 & 0.933 \\
 1 & 2 & 0 & 0 & 0.234 &  0.0002 & 0.027 & -0.0838 & 0.024 & -0.0002 & 0.024 & 0.931 & 0.927 \\
 1 & 2 & 0 & 1 & 0.262 & -0.0003 & 0.030 & -0.0850 & 0.028 &  0.0000 & 0.026 & 0.930 & 0.928 \\
 2 & 1 & 1 & 0 & 0.541 &  0.0000 & 0.033 &  0.0066 & 0.035 & -0.0002 & 0.030 & 0.953 & 0.952 \\
 2 & 1 & 1 & 1 & 0.581 & -0.0016 & 0.034 &  0.0117 & 0.035 & -0.0012 & 0.030 & 0.952 & 0.950 \\
 2 & 1 & 0 & 0 & 0.180 &  0.0006 & 0.026 & -0.0284 & 0.022 &  0.0008 & 0.022 & 0.950 & 0.949 \\
 2 & 1 & 0 & 1 & 0.207 &  0.0007 & 0.031 & -0.0297 & 0.026 &  0.0002 & 0.025 & 0.951 & 0.946 \\
 2 & 2 & 1 & 0 & 0.541 & -0.0001 & 0.033 &  0.0073 & 0.035 &  0.0004 & 0.030 & 0.934 & 0.933 \\
 2 & 2 & 1 & 1 & 0.581 & -0.0008 & 0.035 &  0.0126 & 0.036 & -0.0006 & 0.031 & 0.931 & 0.928 \\
 2 & 2 & 0 & 0 & 0.180 &  0.0027 & 0.025 & -0.0281 & 0.021 &  0.0014 & 0.021 & 0.962 & 0.959 \\
 2 & 2 & 0 & 1 & 0.207 &  0.0006 & 0.030 & -0.0305 & 0.026 & -0.0003 & 0.025 & 0.933 & 0.928 \\
			\hline
		\end{tabular}
	\end{center}
	\vspace{-0.5em}
	\footnotesize{Notes: $T_1$, type of baseline variable; $T_2$, type of control variable; $U$, treatment group; $t$, time point; $\bar{Y}$, observed group mean; $\mu$, marginal group mean; $\hat{\mu}^*$, estimated group mean at the mean covariate \eqref{mcov1}; $\hat{\mu}$, marginal group mean based on the method proposed in \eqref{gr_mean}; SD, standard deviation; CP1/CP2, coverage probability of the 95\% confidence interval based on either  (\ref{lbound},\ref{ubound}) or \eqref{CI_log1}, respectively.}	\vspace{-0.5em}	 
\end{table*}

\begin{table*}[!b]	
	\caption{Summary of various conditional group mean estimators for the logistic design case together with relevant prediction intervals. The number of simulations used is $5000.$}
	\label{table:pi_ul_balanced}
	\vspace{-0.5em}
	\begin{center}
		\setlength\tabcolsep{4.0pt} 
		\footnotesize
		\begin{tabular}{cc cc r rr rr rr rr}
			\toprule
			\multicolumn{2}{c}{Type} &\multicolumn{2}{c}{Group}&\multicolumn{1}{c}{$\lambda$}&\multicolumn{2}{c}{$\bar{Y}-\lambda$}&\multicolumn{2}{c}{$\hat{\lambda}^*-\lambda$}&\multicolumn{2}{c}{$\hat{\lambda}-\lambda$}&\multicolumn{2}{c}{$\hat{\lambda}$}\\ 
			$T_1$ & $T_2$ & $U$ & $t$ & Mean & Mean & SD & Mean & SD & Mean & SD & CP1 &CP2 \\ 
			\midrule
 1 & 1 & 1 & 0 & 0.531 & -0.0009 & 0.027 &  0.0181 & 0.040 & -0.0005 & 0.024 & 0.949 & 0.947 \\
 1 & 1 & 1 & 1 & 0.560 &  0.0002 & 0.029 &  0.0356 & 0.040 & -0.0004 & 0.025 & 0.958 & 0.958 \\
 1 & 1 & 0 & 0 & 0.235 &  0.0005 & 0.026 & -0.0859 & 0.024 & -0.0011 & 0.023 & 0.954 & 0.948 \\
 1 & 1 & 0 & 1 & 0.262 & -0.0004 & 0.030 & -0.0869 & 0.028 & -0.0007 & 0.026 & 0.952 & 0.949 \\
 1 & 2 & 1 & 0 & 0.529 & -0.0003 & 0.027 &  0.0190 & 0.041 & -0.0002 & 0.024 & 0.947 & 0.946 \\
 1 & 2 & 1 & 1 & 0.559 & -0.0005 & 0.029 &  0.0367 & 0.041 & -0.0007 & 0.026 & 0.948 & 0.947 \\
 1 & 2 & 0 & 0 & 0.235 &  0.0001 & 0.026 & -0.0921 & 0.024 & -0.0036 & 0.023 & 0.949 & 0.940 \\
 1 & 2 & 0 & 1 & 0.262 & -0.0006 & 0.029 & -0.0935 & 0.028 & -0.0031 & 0.026 & 0.949 & 0.942 \\
 2 & 1 & 1 & 0 & 0.541 & -0.0001 & 0.031 &  0.0066 & 0.033 & -0.0001 & 0.029 & 0.956 & 0.954 \\
 2 & 1 & 1 & 1 & 0.581 & -0.0002 & 0.034 &  0.0135 & 0.035 &  0.0003 & 0.030 & 0.958 & 0.953 \\
 2 & 1 & 0 & 0 & 0.181 &  0.0000 & 0.025 & -0.0300 & 0.022 & -0.0008 & 0.022 & 0.951 & 0.949 \\
 2 & 1 & 0 & 1 & 0.209 &  0.0005 & 0.030 & -0.0306 & 0.025 & -0.0006 & 0.025 & 0.956 & 0.951 \\
 2 & 2 & 1 & 0 & 0.541 & -0.0002 & 0.032 &  0.0082 & 0.035 &  0.0010 & 0.029 & 0.953 & 0.951 \\
 2 & 2 & 1 & 1 & 0.581 &  0.0002 & 0.033 &  0.0162 & 0.036 &  0.0021 & 0.031 & 0.950 & 0.949 \\
 2 & 2 & 0 & 0 & 0.181 &  0.0020 & 0.024 & -0.0349 & 0.021 & -0.0041 & 0.021 & 0.979 & 0.960 \\
 2 & 2 & 0 & 1 & 0.208 & -0.0002 & 0.030 & -0.0378 & 0.026 & -0.0058 & 0.025 & 0.940 & 0.928 \\
			\hline
		\end{tabular}
	\end{center}
	\vspace{-0.5em}
	\footnotesize{Notes: $T_1$, type of baseline variable; $T_2$, type of control variable; $U$, treatment group; $t$, time point; $\bar{Y}$, observed group mean; $\lambda$, conditional group mean; $\hat{\lambda}^*$, predicted group mean at the mean covariate \eqref{mcov2}; $\hat{\lambda}$, predicted group mean based on the method proposed in \eqref{pred.gr.mean}; SD, standard deviation; CP1/CP2, coverage probability of the 95\% prediction interval based on either \eqref{pred_inverse_logistic} or \eqref{dir.pr}, respectively. }  	\vspace{-0.5em}
\end{table*}		
	
\begin{table*}[!t]	
	\caption{Summary of various marginal group mean estimators for the negative binomial design case together with relevant confidence intervals. The number of simulations used is $5000.$ }
	\label{table:ci_nb_unbalanced}
	\vspace{-0.5em}
	\begin{center}
		\setlength\tabcolsep{2.9pt} 
		\footnotesize
		\begin{tabular}{cc cc r rr rr rr rrr}
			\toprule
			\multicolumn{2}{c}{Type} &\multicolumn{2}{c}{Group}&\multicolumn{1}{c}{$\mu$}&\multicolumn{2}{c}{$\bar{Y}-\mu$}&\multicolumn{2}{c}{$\hat{\mu}^*-\mu$}&\multicolumn{2}{c}{$\hat{\mu}-\mu$}&\multicolumn{3}{c}{$\hat{\mu}$}\\ 
			$T_1$ & $T_2$ & $U$ & $t$ & Mean & Mean & SD & Mean & SD & Mean & SD & CP1 &CP2 &CP3\\ 
			\midrule
 1 & 1 & 1 & 0 & 1.665 &  0.0002 & 0.095 & -0.0081 & 0.082 &  0.0006 & 0.083 & 0.945 & 0.945 & 0.945 \\
 1 & 1 & 1 & 1 & 2.484 &  0.0023 & 0.120 & -0.0106 & 0.109 &  0.0024 & 0.109 & 0.951 & 0.952 & 0.951 \\
 1 & 1 & 0 & 0 & 1.234 & -0.0001 & 0.080 & -0.0066 & 0.065 & -0.0001 & 0.066 & 0.951 & 0.953 & 0.952 \\
 1 & 1 & 0 & 1 & 1.840 &  0.0007 & 0.108 & -0.0085 & 0.093 &  0.0011 & 0.093 & 0.953 & 0.952 & 0.952 \\
 1 & 2 & 1 & 0 & 1.665 &  0.0020 & 0.094 & -0.0070 & 0.081 &  0.0020 & 0.082 & 0.949 & 0.950 & 0.949 \\
 1 & 2 & 1 & 1 & 2.484 &  0.0030 & 0.120 & -0.0105 & 0.109 &  0.0028 & 0.110 & 0.951 & 0.950 & 0.952 \\
 1 & 2 & 0 & 0 & 1.234 &  0.0009 & 0.082 & -0.0058 & 0.068 &  0.0008 & 0.069 & 0.940 & 0.940 & 0.941 \\
 1 & 2 & 0 & 1 & 1.841 &  0.0013 & 0.110 & -0.0086 & 0.094 &  0.0012 & 0.094 & 0.948 & 0.948 & 0.946 \\
 2 & 1 & 1 & 0 & 1.660 & -0.0017 & 0.093 & -0.0037 & 0.082 & -0.0003 & 0.082 & 0.948 & 0.947 & 0.948 \\
 2 & 1 & 1 & 1 & 2.476 &  0.0007 & 0.120 & -0.0053 & 0.110 & -0.0003 & 0.110 & 0.947 & 0.945 & 0.947 \\
 2 & 1 & 0 & 0 & 1.230 &  0.0000 & 0.081 & -0.0034 & 0.067 & -0.0009 & 0.067 & 0.944 & 0.944 & 0.943 \\
 2 & 1 & 0 & 1 & 1.834 & -0.0027 & 0.110 & -0.0047 & 0.095 & -0.0010 & 0.095 & 0.948 & 0.945 & 0.947 \\
 2 & 2 & 1 & 0 & 1.660 & -0.0007 & 0.093 & -0.0038 & 0.080 & -0.0004 & 0.080 & 0.956 & 0.955 & 0.956 \\
 2 & 2 & 1 & 1 & 2.476 &  0.0001 & 0.123 & -0.0055 & 0.111 & -0.0004 & 0.111 & 0.944 & 0.943 & 0.945 \\
 2 & 2 & 0 & 0 & 1.230 &  0.0008 & 0.080 & -0.0020 & 0.067 &  0.0005 & 0.067 & 0.947 & 0.948 & 0.947 \\
 2 & 2 & 0 & 1 & 1.834 &  0.0006 & 0.111 & -0.0028 & 0.095 &  0.0010 & 0.095 & 0.948 & 0.946 & 0.948 \\
		\hline
		\end{tabular}
	\end{center}
	\vspace{-0.5em}
	\footnotesize{Notes: $T_1$, type of baseline variable; $T_2$, type of control variable; $U$, treatment group; $t$, time point; $\bar{Y}$, observed group mean; $\mu$, marginal group mean; $\hat{\mu}^*$, estimated group mean at the mean covariate \eqref{mcov1} with $g$ being the log link function; $\hat{\mu}$, marginal group mean based on the method proposed in \eqref{gr_mean}; SD, standard deviation; CP1/CP2/CP3, coverage probability of the 95\% confidence interval based on \eqref{inv.negb} or \eqref{CI_negb1} or \eqref{lognormal.negb}, respectively.}	 \vspace{-0.5em}
\end{table*}	

\begin{table*}[!t]	
	\caption{Summary of various conditional group mean estimators for the negative binomial design case together with relevant prediction intervals. The number of simulations used is $5000.$} 
	\label{table:pi_nb_unbalanced}
	\vspace{-0.5em}
	\begin{center}
		\setlength\tabcolsep{4.0pt} 
		\footnotesize
			\begin{tabular}{cc cc r rr rr rr rr}
				\toprule
				\multicolumn{2}{c}{Type} &\multicolumn{2}{c }{Group}&\multicolumn{1}{c}{$\lambda$}&\multicolumn{2}{c}{$\bar{Y}-\lambda$}&\multicolumn{2}{c}{$\hat{\lambda}^*-\lambda$}&\multicolumn{2}{c}{$\hat{\lambda}-\lambda$}&\multicolumn{2}{c}{$\hat{\lambda}$}\\ 
				$T_1$ & $T_2$ & $U$ & $t$ & Mean & Mean & SD & Mean & SD & Mean & SD & CP1 &CP2 \\ 
				\midrule
 1 & 1 & 1 & 0 & 1.665 & -0.0016 & 0.093 & -0.0347 & 0.082 & -0.0258 & 0.083 & 0.942 & 0.934 \\
 1 & 1 & 1 & 1 & 2.484 & -0.0018 & 0.119 & -0.0491 & 0.109 & -0.0359 & 0.110 & 0.938 & 0.931 \\
 1 & 1 & 0 & 0 & 1.233 & -0.0001 & 0.079 & -0.0260 & 0.066 & -0.0194 & 0.067 & 0.947 & 0.940 \\
 1 & 1 & 0 & 1 & 1.840 & -0.0018 & 0.110 & -0.0378 & 0.094 & -0.0280 & 0.095 & 0.943 & 0.939 \\
 1 & 2 & 1 & 0 & 1.665 & -0.0008 & 0.094 & -0.0206 & 0.082 & -0.0117 & 0.083 & 0.944 & 0.940 \\
 1 & 2 & 1 & 1 & 2.484 &  0.0015 & 0.122 & -0.0303 & 0.110 & -0.0167 & 0.110 & 0.943 & 0.940 \\
 1 & 2 & 0 & 0 & 1.234 &  0.0000 & 0.080 & -0.0160 & 0.065 & -0.0101 & 0.066 & 0.945 & 0.938 \\
 1 & 2 & 0 & 1 & 1.840 & -0.0015 & 0.110 & -0.0233 & 0.094 & -0.0142 & 0.094 & 0.942 & 0.939 \\
 2 & 1 & 1 & 0 & 1.660 &  0.0001 & 0.094 & -0.0275 & 0.083 & -0.0241 & 0.083 & 0.939 & 0.930 \\
 2 & 1 & 1 & 1 & 2.477 &  0.0021 & 0.121 & -0.0365 & 0.111 & -0.0315 & 0.111 & 0.934 & 0.931 \\
 2 & 1 & 0 & 0 & 1.229 &  0.0003 & 0.080 & -0.0211 & 0.067 & -0.0186 & 0.067 & 0.942 & 0.936 \\
 2 & 1 & 0 & 1 & 1.834 &  0.0018 & 0.109 & -0.0282 & 0.095 & -0.0245 & 0.095 & 0.942 & 0.936 \\
 2 & 2 & 1 & 0 & 1.660 & -0.0015 & 0.094 & -0.0159 & 0.082 & -0.0125 & 0.082 & 0.944 & 0.942 \\
 2 & 2 & 1 & 1 & 2.476 &  0.0013 & 0.120 & -0.0215 & 0.109 & -0.0161 & 0.109 & 0.949 & 0.944 \\
 2 & 2 & 0 & 0 & 1.230 & -0.0008 & 0.079 & -0.0122 & 0.065 & -0.0104 & 0.066 & 0.947 & 0.943 \\
 2 & 2 & 0 & 1 & 1.834 & -0.0003 & 0.109 & -0.0161 & 0.095 & -0.0130 & 0.095 & 0.941 & 0.937 \\
		\hline
		\end{tabular}
	\end{center}
	\vspace{-0.5em}
	\footnotesize{Notes: $T_1$, type of baseline variable; $T_2$, type of control variable; $U$, treatment group; $t$, time point; $\bar{Y}$, observed group mean; $\lambda$, conditional group mean; $\hat{\lambda}^*$, estimated group mean at the mean covariate as in \eqref{mcov2} with $g$ being the log link function; $\hat{\lambda}$, conditional group mean based on the method proposed in \eqref{pred.gr.mean}; SD, standard deviation; CP1/CP2, coverage probability of 95\% prediction interval based on \eqref{pred_inverse_negbin} or \eqref{dir.pr}, respectively.}	\vspace{-0.5em}
\end{table*}			

\begin{table*}[!t]	
	\caption{GLMM Estimation of group mean for hypoglycemia rate and incidence.}
	\label{table:real_glmm}
	\vspace{-0.5em}
	\begin{center}
		\setlength\tabcolsep{3.2pt} 
		\footnotesize
		\begin{tabular}{l rr rr rr rr rr}
			\toprule
			\multicolumn{1}{c}{ } &\multicolumn{2}{c}{Group}&\multicolumn{2}{c}{$\bar{Y}$}&\multicolumn{2}{c}{$\hat{\mu}^*$}&\multicolumn{2}{c}{$\hat{\lambda}$}&\multicolumn{2}{c}{$\hat{\mu}$}\\ 
			Variable & Therapy & Week & Mean & SE & Mean & SE & Mean & SE & Mean & SE\\ 
			\midrule
         Total & BBT &  1-12 & 3.350 & 0.239 & 2.130 & 0.084 & 3.242 & 0.141 & 3.608 & 0.139 \\
  hypoglycemia &  LM &  1-12 & 3.963 & 0.266 & 2.293 & 0.083 & 3.479 & 0.156 & 3.994 & 0.153 \\
          rate & BBT & 13-24 & 5.243 & 0.312 & 3.024 & 0.083 & 4.785 & 0.212 & 5.126 & 0.212 \\
 (per 30 days) &  LM & 13-24 & 5.120 & 0.293 & 3.313 & 0.083 & 5.250 & 0.242 & 5.826 & 0.243 \\  
     Nocturnal & BBT &  1-12 & 0.402 & 0.055 & 0.213 & 0.128 & 0.372 & 0.027 & 0.416 & 0.027 \\
  hypoglycemia &  LM &  1-12 & 0.339 & 0.040 & 0.182 & 0.129 & 0.292 & 0.018 & 0.341 & 0.014 \\
          rate & BBT & 13-24 & 0.651 & 0.067 & 0.336 & 0.123 & 0.593 & 0.046 & 0.657 & 0.045 \\
 (per 30 days) &  LM & 13-24 & 0.509 & 0.052 & 0.291 & 0.125 & 0.482 & 0.031 & 0.551 & 0.024 \\ 
 Proportion of & BBT &  1-12 & 0.712 & 0.024 & 0.826 & 0.219 & 0.724 & 0.012 & 0.720 & 0.004 \\
     patients  &  LM &  1-12 & 0.733 & 0.023 & 0.854 & 0.232 & 0.749 & 0.012 & 0.743 & 0.004 \\
       with $\ge 0$ total & BBT & 13-24 & 0.841 & 0.020 & 0.931 & 0.250 & 0.861 & 0.009 & 0.839 & 0.002 \\
  hypoglycemia &  LM & 13-24 & 0.857 & 0.020 & 0.945 & 0.268 & 0.872 & 0.009 & 0.855 & 0.003 \\
 	Proportion of & BBT &  1-12 & 0.243 & 0.023 & 0.202 & 0.159 & 0.244 & 0.009 & 0.248 & 0.004 \\
 patients with &  LM &  1-12 & 0.250 & 0.023 & 0.172 & 0.163 & 0.213 & 0.008 & 0.222 & 0.005 \\
  $\ge 0$ nocturnal & BBT & 13-24 & 0.402 & 0.027 & 0.342 & 0.151 & 0.376 & 0.011 & 0.369 & 0.004 \\
  hypoglycemia &  LM & 13-24 & 0.330 & 0.026 & 0.303 & 0.153 & 0.343 & 0.010 & 0.339 & 0.006 \\           
			\hline
		\end{tabular}
	\end{center}
	\vspace{-0.5em}
    \footnotesize{Notes: $\bar{Y}$, observed group mean; $\hat{\mu}^*$, estimated group mean at the mean covariate; SE, standard error;  $\hat{\lambda}$, conditional group mean based on the method proposed in \eqref{pred.gr.mean}, $\hat{\mu}$: marginal group mean based on the method proposed in \eqref{gr_mean}.} 	\vspace{-0.5em}
\end{table*}

\end{document}